\documentclass{jfm}

\usepackage{natbib}
\usepackage{graphicx}
\usepackage{hyperref}
\usepackage{amsmath}

\newcommand{\vect}[1]{\boldsymbol{#1}}

\title{Drag reduction on a transonic airfoil}

\author[Quadrio, Chiarini, Banchetti, Gatti, Memmolo \& Pirozzoli]
{Maurizio Quadrio$^1$, Alessandro Chiarini$^1$, Jacopo Banchetti$^1$, Davide Gatti$^2$, Antonio Memmolo$^3$, Sergio Pirozzoli$^4$}

\affiliation{
$^1$ Dipartimento di Scienze e Tecnologie Aerospaziali, Politecnico di Milano, via La Masa 34, 20156 Milano, Italy \\ [\affilskip]
$^2$ Institute for Fluid Mechanics, Karlsruhe Institute of Technology, Kaiserstr. 10, 76131 Karlsruhe, Germany \\ [\affilskip]
$^3$ High Performance Computing Department, CINECA-Interuniv. Cons., 40033 Bologna, Italy\\ [\affilskip]
$^4$ Dipartimento di Ingegneria Meccanica e Aerospaziale, La Sapienza Univ. Roma, Via Eudossiana, Roma, Italy\\ [\affilskip]
}
\begin{document}
\maketitle

\begin{abstract}
Flow control for turbulent skin-friction drag reduction is applied to a transonic airfoil to improve its aerodynamic performance. The study is based on direct numerical simulations (with up to 1.8 billions cells) of the compressible turbulent flow around a supercritical airfoil, at Reynolds and Mach numbers of $Re_\infty= 3 \times 10^5$ and $M_\infty =0.7$. Control via spanwise forcing is applied over a fraction of the suction side of the airfoil. Besides locally reducing friction, the control modifies the shock wave and significantly improves the aerodynamic efficiency of the airfoil by increasing lift and decreasing drag. Hence, the airfoil can achieve the required lift at a lower angle of attack and with a lower drag. Estimates at the aircraft level indicate that substantial savings are possible; when control is active, its energy cost becomes negligible thanks to the small application area. We suggest that skin-friction drag reduction should be considered not only as a goal, but also as a tool to improve the global aerodynamics of complex flows.
\end{abstract}


%%%%%%%%%%%%%%%%%%%%%%%%%%%%%%%%%%%%%%%%%%%%%%%
\section{Introduction}
\label{sec:introduction}
The importance of flow control for the reduction of turbulent skin-friction drag is steadily growing over the years, because of efficiency and environmental reasons. Unfortunately, friction drag reduction of both passive (e.g. riblets) and active techniques are proportional to the fraction of the surface covered by the drag-reducing device. Moreover, in parallel flows, where most research for skin-friction drag reduction has taken place, drag is entirely due to friction. However, the practical appeal of drag reduction is limited in duct flows, where energetic efficiency can be trivially improved by enlarging the cross section of the duct, yielding reduced energy consumption easily and at a small capital cost. 

In more complex flows, the aerodynamic drag includes additional contributions besides the viscous friction, such as pressure drag, parasitic drag, lift-induced drag and wave drag; what ultimately matters is reducing the overall drag. The research community is beginning to explore how skin-friction reduction affects the other drag components. \cite{banchetti-luchini-quadrio-2020} applied spanwise forcing via streamwise-travelling waves to a channel flow with a wall-mounted bump that generates pressure drag. They found that a distributed reduction of friction favourably modifies the pressure field, improving the net energetic benefits by about 50\%. Similarly, \cite{nguyen-ricco-pironti-2021} applied a temporally spanwise-oscillating pressure gradient to a channel flow with transverse bars at the wall, and found that pressure drag is reduced as nearly as friction drag, although the overall net energy budget remains slightly negative. 

One of the applications where drag reduction entails obvious benefits is the aircraft, where achieving aerodynamic efficiency is the key. Flow over aircraft wings features pressure gradients and shock waves, which are responsible for significant drag penalty, amenable to control through a variety of techniques \citep{bushnell-2004}, including placement of small control bumps to modify the shock \citep{bruce-collis-2015}, and use of oscillatory blowing to delay flow separation \citep{seifert-etal-1993}. Most of these studies are experimental, or based on Reynolds Averaged Navier--Stokes (RANS) simulations. Recently, \cite{atzori-etal-2020} studied with high-fidelity Large Eddy Simulations (LES) the effect of uniform blowing or suction on the incompressible flow past a NACA4412 airfoil at Reynolds number (based on free-stream velocity and chord length) of $Re=200,000$. They found that the wing efficiency improves up to 11\% when uniform suction is applied to the suction side, leading to friction drag increase, but pressure drag reduction. \cite{fahland-etal-2021} demonstrated the potential of blowing on the pressure side under various conditions, achieving a maximum total net drag saving of 14\%. \cite{kornilov-2021} carried out an experimental study of blowing/suction on two-dimensional low-speed airfoils, and provided ideal estimates of the power spent for actuation. \cite{albers-schroeder-2021} studied with implicit LES the same airfoil considered by \cite{atzori-etal-2020}, but controlled the flow via spanwise-travelling waves of wall-normal deformation. They generalised their previous results based on a different wing section \citep{albers-meysonnat-schroeder-2019} and demonstrated that control alters both friction and pressure drag, thus improving the overall aerodynamic performance of the wing. 

All these works have considered the flow in incompressible or subsonic regimes. However, there are reasons to suspect \citep[see e.g.][in the context of riblets]{mele-tognaccini-catalano-2016} that a further advantage of reducing skin friction on a wing resides in the ability to interact with the position and strength of the shock waves which may form in the transonic regime. In this work we present the first direct numerical simulation (DNS) of the turbulent flow over an airfoil in the transonic regime, where flow control originally targeted to friction reduction is applied. Specifically, we explore to what extent a localised control for skin-friction reduction interacts with the shock and alters the aerodynamic performance of the airfoil. The results are also extrapolated to the whole aircraft. The active control technique chosen for the study is the streamwise-traveling waves of spanwise forcing \citep{quadrio-ricco-viotti-2009}, which produce large (hence easily measurable) effects and large net savings as well. Although physical actuators to implement such forcing on an airplane are currently unavailable, the general conclusions are expected to be valid for any skin-friction reduction technology, be it active or passive.

%%%%%%%%%%%%%%%%%%%%%%%%%%%%%%%%%%%%%%%
\section{Methods}
\label{sec:methods}

\begin{figure}
\includegraphics{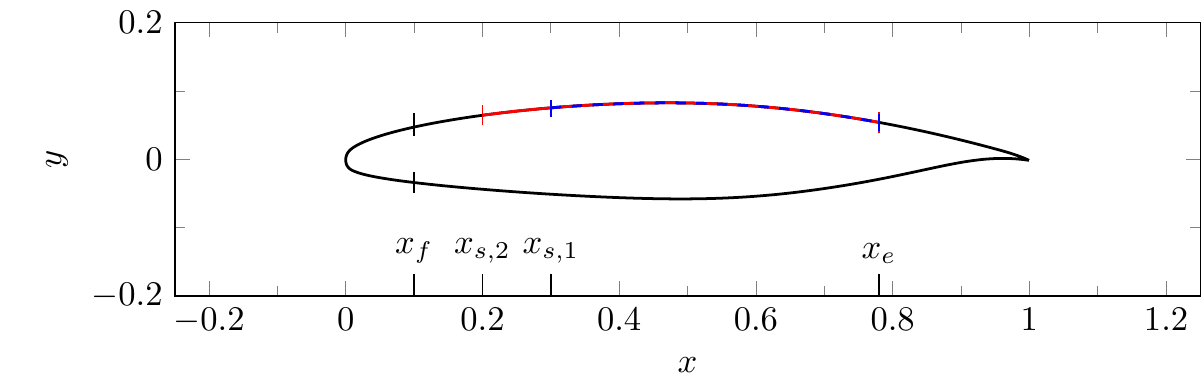}
\caption{Geometry of the V2C airfoil. Forcing is applied at $x_f$ (on both sides) to initiate transition. $x_s$ and $x_e$ denote start and end of the suction-side actuated region for cases C1 (blue dashed line) and C2 (red continuous line).}
\label{fig:geometry}
\end{figure}

We study by DNS the transonic flow around the supercritical V2C airfoil (see figure \ref{fig:geometry}), designed by Dassault Aviation within the European research program TFAST. Reynolds and Mach numbers are set to $Re_\infty = U_\infty c /\nu_\infty= 3 \times 10^5$ and $M_\infty=U_\infty/a_\infty=0.7$, where $c$ is the airfoil chord and $U_\infty$, $\nu_\infty$ and $a_\infty$ are the free-stream velocity, kinematic viscosity and sound speed. $U_\infty$ and $c$ are the reference velocity and length, unless otherwise noted. The $x$, $y$ and $z$ axes indicate the chord-wise, vertical and spanwise directions. The angle of attack is $\alpha=4^{\circ}$, which corresponds to the maximum aerodynamic efficiency of the profile. 

The DNS code \citep[see][for a detailed description]{memmolo-bernardini-pirozzoli-2018} solves the compressible Navier--Stokes equations for a calorically perfect gas. It is based on a baseline second-order, energy-consistent finite-volume discretization, with local activation of a third-order shock-capturing WENO numerical flux, controlled by a modified Ducros sensor \citep{ducros-etal-1999}. Time advancement uses a low-storage, third-order Runge--Kutta scheme. At the far field, characteristics-based non-reflective boundary conditions are used \citep{poinsot-lele-1992}, whereas periodicity is enforced in the spanwise direction. Discretisation is based on a C-type mesh, with radius of $25c$; the outflow is placed at $25c$ from the trailing edge. The domain extends for $0.1c$ in the spanwise $z$ direction, to ensure decorrelation of all the flow structures on the airfoil and in the wake \citep{zhang-samtaney-2016,hosseini-etal-2016}. The incoming flow is laminar. As done by \cite{schlatter-orlu-2012}, transition to turbulence is enforced on both sides of the airfoil via a time-varying wall-normal body force located around $x = x_f = 0.1c$, in such a way as to minimize disturbances associated with boundary layer tripping.

Streamwise-travelling waves of spanwise velocity are applied to a portion of the suction side of the wing, according to 
\[
w_w (x,t) = A f(x) \sin \left( \kappa_x x - \omega t \right)
\]
where $A$ is the maximum forcing amplitude and $\kappa_x$ and $\omega$ are the spatial and temporal frequencies of the wave. As in \cite{yudhistira-skote-2011}, a smoothing function $f(x)$ is used to raise the spanwise velocity at the initial position $x_s$ and then return it to zero at $x_e$.

Two forcing configurations are considered, hereinafter referred to as cases C1 and C2: they have been selected after a preliminary study, such that the stronger forcing scheme C2 yields flow separation past the shock wave, whereas the milder forcing scheme C1 barely does so. In both cases the actuated region starts past the tripping zone, and ends past the shock. In particular, case C1 has $x_s=0.3$, $x_e=0.78$, $A=0.5$, $\omega=11.3$ and $\kappa_x=161$. In C2 the actuated region starts earlier at $x_s=0.2$, and the forcing amplitude is larger, i.e. $A=0.684$. For the C2 forcing, this corresponds to $A^+ \approx 6.6$, $\omega^+ \approx 0.06$ and $\kappa_x^+ \approx 0.013$ after expressing quantities in viscous units computed with the average value of the friction velocity along the actuated region: this is not far from the incompressible channel flow maximum net saving, yielding about 33\% drag reduction and 20\% net power savings at a friction Reynolds number of $Re_\tau =200$. (Note that forcing optimization on the wing would by itself deserve a more detailed investigation.)

Six DNS have been carried out. Four employ a baseline grid with 536 million cells and include a reference uncontrolled case, as well as C1 and C2 (the latter repeated at a different angle of attack, see later \S\ref{sec:forces}). This mesh size is comparable to that used by \cite{zauner-detullio-sandham-2019} for the same profile but at the larger $Re_\infty=5 \times 10^5$. The uncontrolled and C2 cases have also been computed on a finer mesh made by 1.8 billions cells, to check grid convergence. The baseline mesh has $4096 \times 512 \times 256$ cells, with uniform spacing in the spanwise direction and a hyperbolic-tangent clustering in the wall-normal direction to achieve sufficient resolution close to the airfoil and in the wake. The finer mesh, including $6144 \times 768 \times 384$ cells, has the number of cells increased by 50\% in each coordinate direction. An {\em a posteriori} check has confirmed that requirements for a fully-resolved DNS \citep{hosseini-etal-2016}, namely $\Delta x^+ < 10, \Delta y^+ < 0.5, \Delta z^+ < 5$ are satisfied in the near-wall region. The same authors determined that for such flow this resolution is also adequate to resolve the wake region.
The simulations are advanced in time with a constant step, selected to maintain the maximum Courant--Friedrichs--Lewy number below unity. Specifically, we use $\Delta t= 1.5 \times 10^{-4}$ for the baseline mesh, and $\Delta t = 1 \times 10^{-4}$ for the fine mesh. The flow statistics are accumulated over a time interval of $40 c/U_\infty$, after reaching statistical equilibrium.

%%%%%%%%%%%%%%%%%%%%%%%%%%%%%%%%%
\section{Results}
\label{sec:results}

\subsection{Instantaneous and mean fields}

\begin{figure}
\centering
\includegraphics[width=0.85\textwidth]{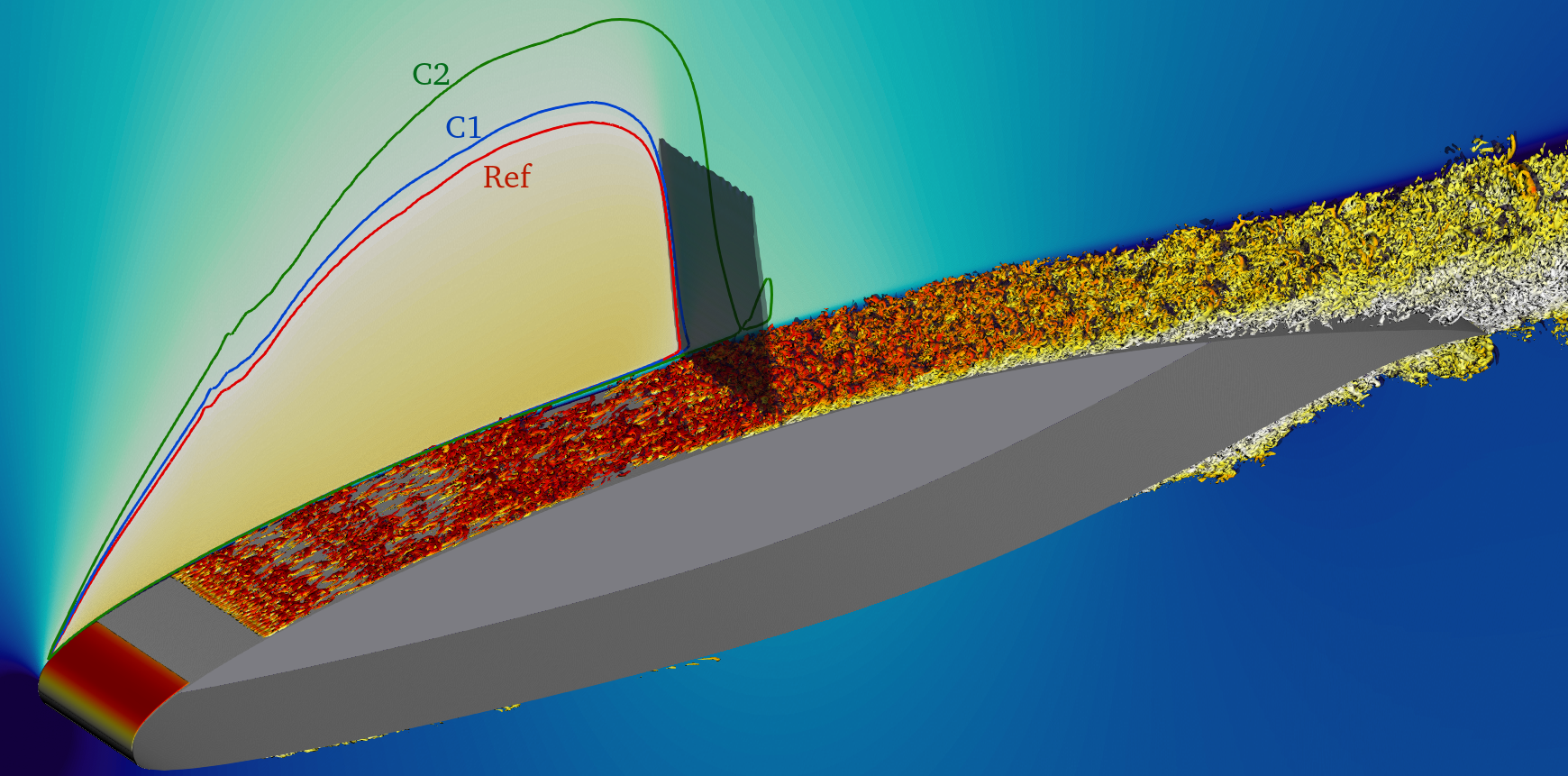}
\caption{Instantaneous turbulent structures (uncontrolled case) visualised in terms of iso-surfaces of the swirling strength ($\lambda_{ci}=100$), and coloured with the turbulence kinetic energy $k$ (white-to-red colourmap for $0 \le k \le 1$). The background colourmap is for the mean Mach number (symmetric blue-to-red color map for $0.5 \le M \le 1.5$). Sonic lines at $M=1$ are drawn for the uncontrolled (red), C1 (blue) and C2 (green) flow cases. The shock wave is shown via the grey isosurface of $\partial \rho / \partial x = 10$.}
\label{fig:structures-shock}
\end{figure}

An overview of the mean and instantaneous fields in the uncontrolled case is provided by figure \ref{fig:structures-shock}, where instantaneous vortical structures are visualised via iso-surfaces of the imaginary part $\lambda_{ci}$ of the complex conjugate eigenvalue pair of the velocity gradient tensor \citep{zhou-etal-1999}. The shock is also shown, together with the mean Mach number in the background colourmap. The three sonic lines at $M=1$ are shown for the reference, C1 and C2 cases. 
The flow becomes supersonic at the nose and remains laminar up to the tripping. The supersonic region extends up to $x \approx 0.5c$, where the shock wave produces an abrupt recompression. Flow control shifts the shock downstream, enlarging the supersonic region, whose streamwise and vertical dimensions increase from $D_x=0.47c$, $D_y=0.35c$ (reference) to $D_x=0.48c, D_y=0.36c$ (C1), and $D_x=0.52c, D_y=0.42c$ (C2). The shock strength correspondingly increases, with the pressure jump measured at $y=0.2$ increasing from $\Delta p=0.121$ (reference) to $\Delta p=0.136$ (C1), to $\Delta p=0.167$ (C2). At the same time, the maximum Mach number increases from $M=1.087$ (reference) to $M=1.093$ (C1), to $M=1.116$ (C2), whereas its position is nearly unaffected, at $(x,y)\approx (0.39c,0.094c)$. These modifications are consistent with decreased friction over the actuated region, yielding increased supersonic speeds.

\begin{figure}
\centering
\includegraphics[width=\textwidth]{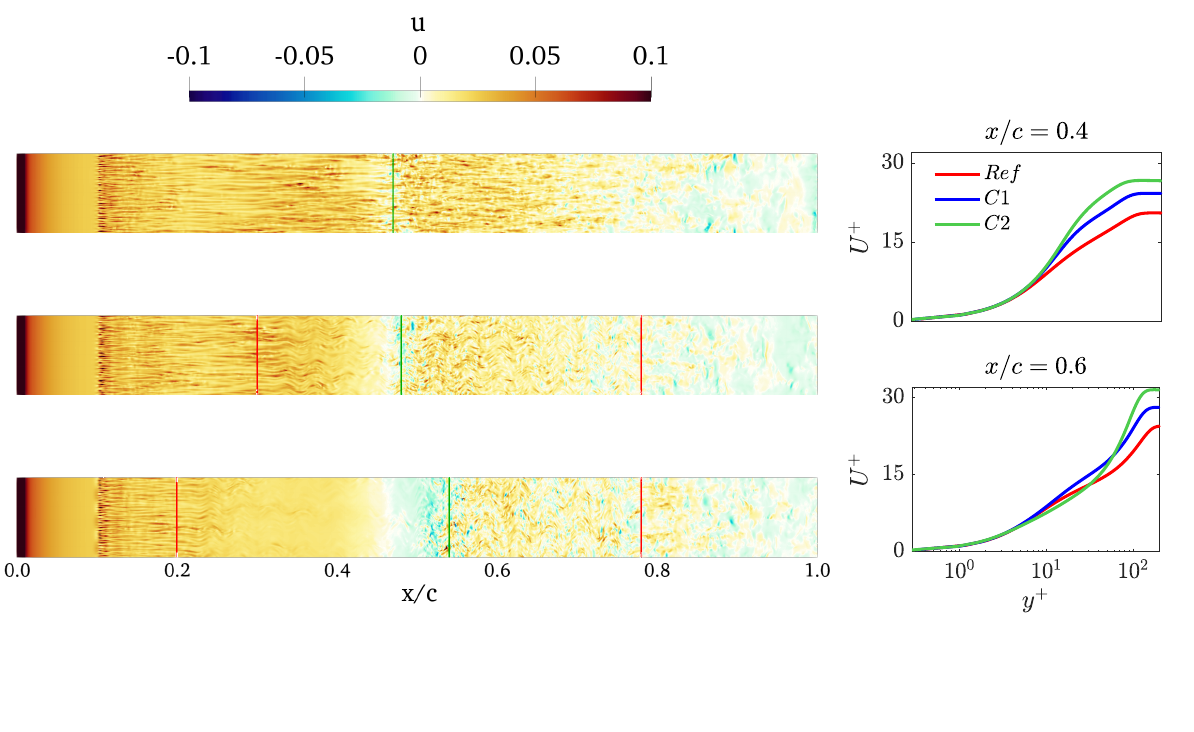}
\caption{Instantaneous chord-wise velocity at the first grid point off the wing surface over the suction side, for uncontrolled (top), C1 (middle) and C2 (bottom) flow cases. Red lines mark the boundaries of the actuated region, and green lines the position of the shock. The panels on the right plot the mean velocity profile in local wall units upstream and downstream of the shock.}
\label{fig:uy1}
\end{figure}
The development of the near-wall flow along the suction side is visualised in figure \ref{fig:uy1} for the three flow cases, where the instantaneous chord-wise velocity (a proxy for the instantaneous wall friction) is shown at the first grid point off the wall. The right panels plot the mean velocity profile at $x/c=0.4$ and $x/c=0.6$, i.e. before and after the shock. The boundary layer is laminar up to the tripping location, whence a pattern of alternating low- and high-speed streaks is generated. Then, in the uncontrolled case the fluctuations undergo transient decay, and then grow further up to $x \approx 0.46c$, where interaction with the shock wave disrupts the streaks. Immediately past $x_s$, control produces visible spanwise meandering in the developing streaks and affects the transition process, so that the streaks nearly vanish at $x_s + 0.1c$. Past the shock, the uncontrolled case features scattered spots of backflow $u<0$, that are most intense in the C2 flow case, suggesting separation of the boundary layer in mean sense.

%--------------------------------------
\subsection{Wall friction and pressure}

\begin{figure}
\centering
\includegraphics[width=0.9\textwidth]{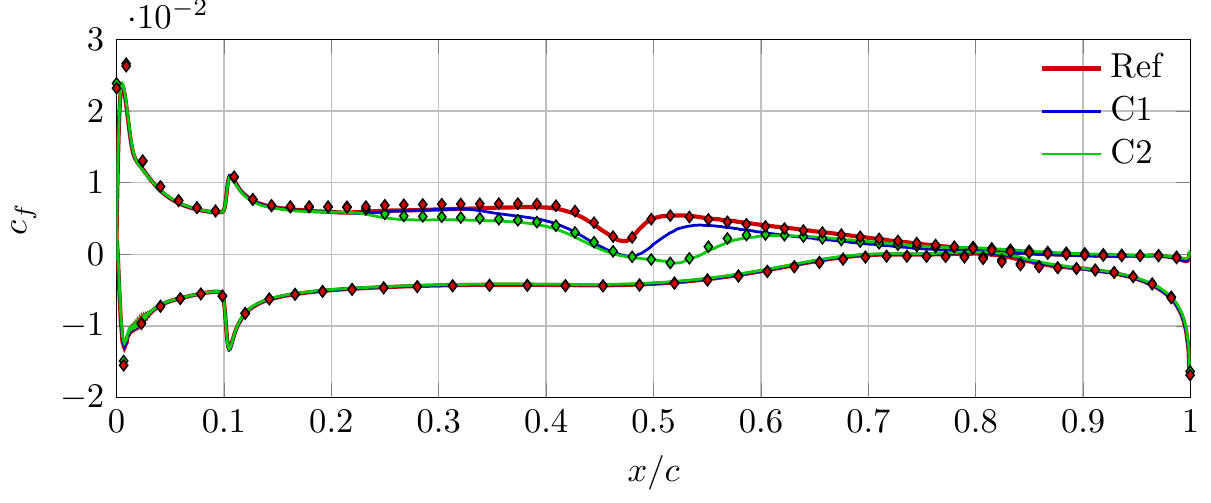}
\includegraphics[width=0.9\textwidth]{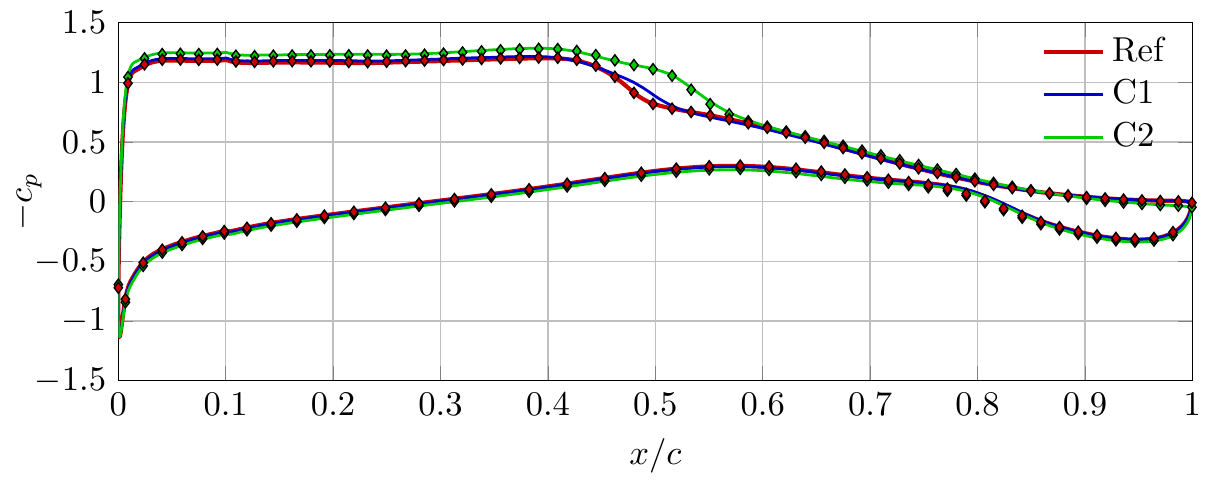}
\caption{Mean friction coefficient $c_f$ (top) and mean pressure coefficient $c_p$ (bottom). Symbols denote the uncontrolled and C2 cases computed on the finer mesh. Note that green and red symbols almost overlap on the pressure side.}
\label{fig:cp-cf-coeff}
\end{figure}

Figure \ref{fig:cp-cf-coeff} plots the mean friction and pressure coefficients $c_f=2 \tau_w/(\rho_\infty U_\infty^2)$ and $c_p=2(p_w-p_\infty)/(\rho_\infty U_\infty^2)$. $\tau_w= \mu \hat{\vect{t}} \cdot \partial \vect{u} / \partial n$ is the wall shear stress, with $\hat{\vect{t}}$ the tangential unit vector and $\partial / \partial n$ the derivative in the wall-normal direction. $\mu$, $\rho$ and $p_w$ are the dynamic viscosity, density and wall pressure. 

Results from the baselines mesh (solid lines) and the finer mesh (symbols) are overlapping, and especially $c_p$ is very nearly mesh independent. Only the peak of $c_f$, in the relatively unimportant nose region of the airfoil is slightly under-resolved by the baseline mesh. Changes between uncontrolled and actuated cases are, however, very well predicted on both meshes. 
Since the pressure side is not actuated and the flow properties are virtually unaffected, only the suction side is considered hereafter. Past the leading-edge peak, $c_f$ decreases rapidly in the laminar region, up to $x_f$, where the effect of numerical tripping is clearly visible. Further downstream, $c_f$ drops because of the shock-wave/boundary-layer interaction at $x \approx 0.45c$ and, after a transient increase, it slowly drops again, eventually becoming negative right upstream of the trailing edge. In the uncontrolled case, despite the negative $u$ fluctuations observed in the instantaneous field in figure \ref{fig:uy1}, $c_f$ past the shock wave remains positive. 
In the controlled cases, forcing does its job of reducing friction in the actuated part of the wing. It is known \citep{quadrio-ricco-2004,skote-2012} that some spatial extent is required for drag reduction to develop. Once this is accounted for, the local skin friction reduction is in line with expectations based on incompressible channel flow simulations. In both controlled cases $c_f$ becomes negative past the shock wave, but negligibly so for C1, in agreement with the instantaneous visualisations of figure \ref{fig:uy1}. 
% A final remark concerns the position of the local minimum of $c_f$ immediately after the shock wave. Since the shock wave moves downstream with the control, the local $c_f$ minimum moves downstream too: it is found at $x =0.47$, $0.48$ and $0.51$ for the no-control, C1 and C2 cases.

The pressure coefficient, after the leading-edge expansion, features a plateau which extends all the way to the near-shock region, as per design of the airfoil. Then, in the rear part $c_p$ progressively increases, and becomes nearly zero at the trailing edge. In the controlled cases two distinct effects are observed: the compression associated with the shock wave is shifted downstream, and the leading-edge expansion intensifies. The recirculating region in the controlled cases mitigates the adverse pressure gradient near the shock, yielding a milder slope of the $c_p(x)$ curve. Hence the shock wave moves downstream, leading to a larger supersonic region with a higher velocity therein, and therefore to a stronger expansion in the fore part of the airfoil. Both effects are more evident in the C2 case, designed to yield stronger effects. Overall, the control modifies the $c_p$ distribution in a way that is consistent with a slight increase of the free-stream Mach number, but only on the suction side.

%-------------------------------------------------------------
\subsection{Aerodynamic forces and aircraft-level performance}
\label{sec:forces}

\begin{table}
\centering
\begin{tabular}{c| c c c c c c c c}
          & Uncontrolled    & C1     &  $\Delta_1$    & C2    & $\Delta_2$ & & C2 ($\alpha=3.45^\circ$) & $\Delta_2$ \\
\hline
$C_l$     & 0.740  & 0.751   &  +1.5\%        & 0.825  & +11.3\%    & &  0.730                    & -1.3\% \\
$C_d$     & 0.0247 & 0.0236  &  -4.5\%        & 0.0245 & -0.8\%     & &  0.0210                   & -15.0\% \\
$C_{d,f}$ & 0.0082 & 0.0076  &  -7.3\%        & 0.0071 & -13.4\%    & &  0.0074                   & -9.7\% \\
$C_{d,p}$ & 0.0165 & 0.0161  &  -2.4\%        & 0.0174 & +5.5\%     & &  0.0136                   & -17.6 \\
$C_l/C_d$  & 29.7   & 31.7    &  +6.8\%        & 33.7   & +13.5\%    & &  34.8                    & +17.2\% \\
\end{tabular}
\caption{Lift and drag coefficients ($C_l$, $C_d$) of the airfoil, and splitting of drag coefficient into friction and pressure contributions ($C_{d,f}$, $C_{d,p}$), for uncontrolled, C1 and C2 flow cases. $\Delta$ stands for relative change, and the last two columns refer to the C2 case computed for an angle of attack $\alpha=3.45^\circ$ (see text).}
\label{tab:forces}
\end{table}

The control-induced modifications to friction and pressure favourably affect lift and drag. Table~\ref{tab:forces} compares the lift and drag coefficients $C_l$ and $C_d$ of the airfoil for the uncontrolled and controlled cases. Friction and pressure contributions to the overall drag are separately accounted for, and reported as $C_{d,f}$ and $C_{d,p}$. Control reduces the friction drag by 7.3\% and 13.4\% for the C1 and C2 cases, respectively. These reductions are quite substantial, as control is applied only to about one quarter of the airfoil surface. Naturally, friction drag reduction is larger for the C2 case, owing to its longer actuation region and stronger intensity. Pressure drag changes are instead quite different in the two cases: $C_{d,p}$ decreases by 2.4\% in the C1 case, and increases by 5.5\% in the C2 case. These combined changes result into reduction of the total drag in both cases, quantified in 4.5\% for C1, and in a marginal 0.8\% for C2. However, an additional crucial change is the increase of the lift coefficient. In agreement with the changes in the pressure distribution shown in figure \ref{fig:cp-cf-coeff}, the increase of $C_l$ is minimal for C1 (+1.5\% only), but quite large for C2 (+11.3\%). The wing efficiency, therefore, is significantly enhanced in both cases, by 6.8\% for C1, and by 13.5\%  the C2.

Increasing the wing efficiency implies that the lift required to balance the aircraft weight can be obtained at lower angle of attack, hence with lower drag penalty. To determine this contribution to drag reduction, we start by computing $C_l-\alpha$ and $C_d-\alpha$ maps (not shown) for the uncontrolled airfoil via auxiliary RANS simulations, using a modified version of the same flow solver with the Spalart--Allmaras turbulence model. Under the assumption that small changes of $\alpha$ do not alter the control-induced percentage changes of the aerodynamic forces, the new angle of attack is identified: for the C2 case, it is $\alpha=3.45^\circ$, for which an additional DNS is carried out (its results are shown in the last two columns of table \ref{tab:forces}). The lift coefficient is $C_l=0.730$, hence slightly less than expected, but the drag coefficient drops to $C_d = 0.0210$, i.e. with 15\% of drag reduction.

It is instructive to tentatively scale these figures, although with inevitable approximations, up to the full aircraft. As an example, we consider the wing-body configuration DLR-F6 defined in the Second AIAA CFD drag prediction workshop \citep{laflin-etal-2005}, with flight conditions $M_\infty=0.75$ and $Re_\infty= 3 \times 10^6$. The reference lift coefficient is $C_L=0.5$, obtained at angle of attack $\alpha=0.52^{\circ}$, at the cost of $C_D=0.0295$. We look for the achievable drag reduction when control C2 is applied. In doing this, the following simplifying assumptions are made: (i) the wing is responsible for the entire lift and its non lift-induced drag is 1/3 of the total drag; (ii) changes to $C_l$ and $C_d$ resulting from control are constant along the wing span, and do not change with $\alpha$, $M_\infty$ and $Re_\infty$, so that the values reported in Table \ref{tab:forces} apply.
Using information available from {\tt https://aiaa-dpw.larc.nasa.gov/Workshop2/DPW\_forces\_WB\_375}, applying C2 control would reduce the angle of attack to $\alpha=0.0125^{\circ}$, thus yielding $C_D =0.0272$, hence with a drag reduction of approximately 8.5\%; the additional small benefit of direct skin-friction reduction leads to about 9\% reduction for the aircraft drag. It is also important to note that the actuation power required by the C2 forcing is very small. Under the assumption of actuation with unitary efficiency, it equals the power transferred to the viscous fluid by the boundary forcing; by computing its time-average value after the DNS, it is measured to be $5.5 \times 10^{-4} \rho_\infty U_\infty^3$. Owing to the localized actuation (the actuated area is approximately one-fourth of the wing surface, and one twelfth of the entire aircraft surface) the actuation power would thus be about 1\% of the overall power expenditure.

%%%%%%%%%%%%%%%%%%%%%%%%%%%%%%%%%%
\section{Concluding discussion}
\label{sec:conclusions}

The first DNS of the controlled compressible turbulent transonic flow over a wing slab at $M_\infty=0.7$ and $Re_\infty=3 \times 10^5$ has been presented. The aerodynamic performance of the wing is improved by using active spanwise wall forcing to locally reduce skin friction over a portion of the wing suction side. The forcing causes stronger expansion in the fore part and a delayed, more intense shock. This is equivalent to an increase of the Mach number on the suction side of the wing, and significantly improves the lift/drag ratio. For a constant angle of attack, the aerodynamic efficiency has been observed to increase by 13.5\% (with drag decreased by 0.8\% only). Higher aerodynamic efficiency allows the required lift to be achieved at a lower angle of attack, yielding significant reduction of the total drag, which we have quantified via DNS to be about 15\%. We have also estimated that this may lead to overall drag reduction of about 9\% for a full aircraft in cruise flight, and that the energy cost for the active control is about $1\%$ of the total power expenditure. Realizing that a local friction control may yield global benefits has enormous importance in terms of both practical feasibility and cost/benefit assessment.

We close this paper acknowledging possible limitations. First, the $Re_\infty$ value considered here is clearly not representative of an airplane in transonic flight, and serious design attempts should consider higher $Re$. Luckily, we know \citep{gatti-quadrio-2016} that spanwise forcing and, more generally, skin-friction drag reduction techniques retain their effectiveness at higher $Re$, although the optimal wall actuation parameters may change, even if viscous scaling is assumed. Furthermore, the reported drag reduction of 9\% should by no means taken as a maximum achievable gain. Indeed, the available forcing information for incompressible flow over a plane wall does not easily translate to aircraft in cruise conditions; moreover, the plane-channel maximum power saving condition is most probably not optimal here. Locating where, when and how much the control should be activated is a new optimization problem, whose solution might show much better performance. Above all, one should always be aware of the great challenge of designing actuators capable to efficiently meet the required specifications. The general ideas discussed here, however, retain their validity also for other schemes for friction reduction, including passive strategies, e.g. riblets, that remain the most obvious choice for applications.

All in all, we believe that considering skin-friction drag reduction as a tool and not only as a goal in flows where friction drag is not the key target for optimisation might open new avenues for a more widespread use of flow control.
 
\section*{Acknowledgments}
Computational resources have been provided by CINECA through grant TragFoil and by HLRS through grant TuCoWi. We acknowledge F. Billard and M. Braza for providing data about the V2C airfoil.

\section*{Funding} 
This research received no specific grant from any funding agency, commercial or not-for-profit sectors.

\section*{Declaration of Interests} 
The authors report no conflict of interest.

\bibliographystyle{jfm}

\end{document}